\documentclass[journal]{IEEEtran}
\ifCLASSINFOpdf
   \usepackage{graphicx}
\else
  \usepackage[dvips]{graphicx}
\fi
\usepackage[cmex10]{amsmath}

\usepackage[utf8]{inputenc}
\usepackage{tabu}
\usepackage{color}
\usepackage{hyperref}
\usepackage{cleveref}

\begin{document}
%
\title{Non-Verbal Communication Analysis in Victim-Offender Mediations}
%
%
%

\author{Víctor~Ponce-López,
\thanks{V. Ponce-López is with the IN3 at the Open University of Catalonia,
the Dept. MAiA at the University of Barcelona,
and the Computer Vision Center at the Autonomous University of Barcelona%
.} Sergio~Escalera,
\thanks{S. Escalera is with the Dept. MAiA at the University of Barcelona,
and the Computer Vision Center at the Autonomous University of Barcelona%
.} Marc Pérez,\thanks{M. Pérez is a B.S. student in Computer Science and Mathematics at the Faculty of Mathematics. He has collaborated in this work with the Dept. MAiA at the University of Barcelona%
.} Oriol Janés,\thanks{O. Janés obtained his B.S. in Computer Science at the Faculty of Mathematics. He has collaborated in this work with the Dept. MAiA at the University of Barcelona%
.} and~Xavier~Baró
\thanks{X. Baró is with the EIMT at the Open University of Catalonia,
and the Computer Vision Center at the Autonomous University of Barcelona.%
}}%


%
%

\markboth{Under consideration at Pattern Recognition Letters}%
{Shell \MakeLowercase{\textit{et al.}}: Non-Verbal Communication Analysis in Victim-Offender Mediations}
%



\maketitle

\begin{abstract}
In this paper we present a non-invasive ambient intelligence
framework for the semi-automatic analysis of non-verbal
communication applied to the restorative justice field. In
particular, we propose the use of computer vision and social
signal processing technologies in real scenarios of
Victim-Offender Mediations, applying feature extraction techniques
to multi-modal audio-RGB-depth data. We compute a set of
behavioral indicators that define communicative cues from the
fields of psychology and observational methodology. We test our
methodology on data captured in real world Victim-Offender
Mediation sessions in Catalonia in collaboration with
the regional government. We define the ground truth based on expert opinions 
when annotating the observed social responses.
Using different state-of-the-art binary classification
approaches, our system achieves recognition accuracies of 86\%
when predicting satisfaction, and 79\% when predicting both
agreement and receptivity. Applying a regression strategy, we
obtain a mean deviation for the predictions between 0.5 and 0.7 in
the range [1-5] for the computed social signals.
\end{abstract}


\IEEEpeerreviewmaketitle

\section{Introduction}

\begin{figure*}[ht]
    \centering
    \includegraphics[width=\textwidth]{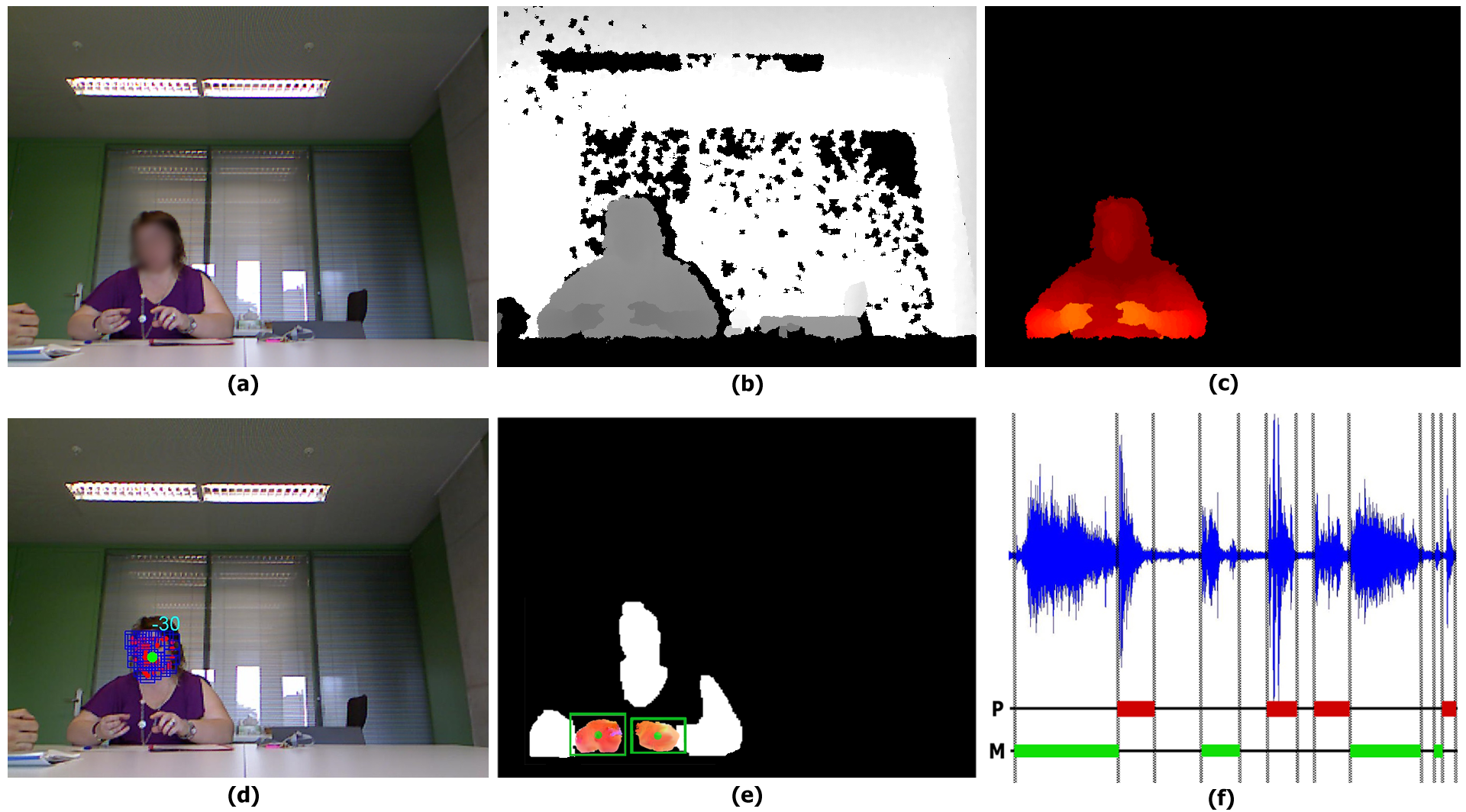}
    \caption{Examples of the multi-modal feature extraction
    . Images (a) and (b) are the RGB and depth images, respectively. Image (c) shows the upper body obtained from the Random Forest user segmentation.
    In image (d), both face detection and head pose estimation are shown. Hand segmentation is shown in image (e).
    Across the regions segmented by color, the optical flow is shown in the regions in which there is greatest movement, identified as being the hands.
    Finally, image (f) illustrates the speaker diarization process with the two participants involved in the VOM session. The participants belong either to a party P or to the mediators M. Clusters belonging to each participant are obtained from the input signal, estimating the speech time of each segment, as well as the speech pauses/interruptions.}
    \label{fig:features} 
\end{figure*}

\IEEEPARstart{R}{ESTORATIVE} justice is an
international social movement for the reform of criminal justice.
This approach to justice focuses on the needs of the victims, who
take an active role in the process, while offenders are encouraged
to take responsibility for their actions \emph{to repair the harm
they have done}~\cite{weitekamp:restorativeJust}. It differs,
therefore, with respect to the classic justice framework, where
justice seeks to satisfy abstract legal principles or to directly
punish offenders. One of the common procedures offered to victims
is the possibility of exchanging their impressions with a
mediator, in a program known as the Victim-Offender Mediation
(henceforth VOM) program. Given the sensitive nature of the cases,
the process consists initially of a set of individual encounters,
where each party involved (i.e., victim or offender) attends an
interview or meeting with a mediator to analyze the problem in
depth. The decision is then taken as to whether the victim and the
offender might engage in a joint encounter.
Figure~\ref{fig:features} (a) shows an example of a real VOM
scenario.


In the VOM process, the goal is to reach a restitution agreement
by seeking to balance the interests of each of the parties,
conditioned of course by the events that have occurred and the
legal proceedings. This agreement can be reached in one of two
ways. First, there are pre-conditioning factors to a case, given
its particular facts, which make mediation feasible or not.
Second, high levels of agreement and expressed satisfaction
between the parties and the mediator are indicators of whether the
VOM process is likely to end in success or
failure~\cite{umbreit:vom}. The emergence of these indicators
depends on a large set of factors that are not only concerned with
the professionalism of the mediator, but are also related to other
factors including the applicability of mediation, the
participants' traits, human relationships, and first impressions,
among others. Furthermore, if we examine each of the participants
(victim, offender, and mediator), certain characteristics,
including their cultural background, education, and social status,
are likely to have a high impact on the success or otherwise of
the process~\cite{pentland:computation,pentland:honestsignals}.
For example, feelings of inferiority may emerge during the
conversations if the mediator employs higher language skills than
those employed by one or other of the parties involved. In such
cases, the mediation process is perhaps more susceptible to
failure than it is to success. Thus, it is important to emphasize
the particular behavioral traits of those involved in a VOM from a
psychological perspective.

Participant roles are clearly defined in these conversational
processes, as they are in similar scenarios, such as job
interviews. The mediator explains the process and listens to the
other parties, maintaining his or her impartiality at all times,
whereas the victim and offender are more concerned with protecting
their own interests and may appear quite wrapped up in the problem
they face. In this sense, no standard guidelines exist for
establishing the best course of actions or identifying the
psychological mechanisms for achieving the desired mediation
goals. Moreover, the mediators have often received a different
type of training focused neither in psychology nor in
observational methodology, two fields concerned with understanding
how biological, psycho-social, and environmental factors can
impact participants' moods. However, if these techniques were
studied and established as self-knowledge tools for both
communication and intervention, mediators could obtain valuable
feedback from the mediations and guide the process towards
success.

In this context, multi-modal intelligent systems can be used 
to achieve these objectives. This information can be
analyzed for different modules in order to extract features from
each source separately. They can then be combined so as to define
and recognize communicative indicators.

\subsection{Related work} \label{sec:relatedwork}

The Restorative Justice approach focuses on the
personal needs of victims. As discussed above, achieving success
in the VOM sessions depends largely on how the participants
communicate with each other. A large number of techniques can be
found in the literature for application in VOM. A good example of
this is provided by Umbreit's handbook~\cite{umbreit:vom}. This resource offers an empirically
grounded, state-of-the-art analysis of the application and impact
of VOM. It provides practical guidance and resources for VOM in
the case of property crimes, minor assaults, and, more recently,
crimes of severe violence, where family members of murder victims
request a meeting with the offender. Since most of these cases
addressed are of a highly sensitive nature, participants are
likely to manifest emotional states, when interacting with the
others, that can be physically observed through their non-verbal
communication~\cite{knapp1997nonverbal}. 

Recently, a number of studies have proposed ways in which
personality traits can be inferred from multimedia
data~\cite{Mohammadi:2013:PRP:2522848.2522857} and which can be
applied directly to the approach taken by Restorative Justice. The
prediction of these responses takes a particular interest in
meetings involving a limited number of participants. For instance,
in \cite{gatica:groupleaders} the goal was both to detect the
social signals produced in small group interactions and to
emphasize their importance markers. In addition, the works of
\cite{gatica:bodycues,Aran:2013:OKI:2522848.2522859} combined
several methodologies to analyze non-verbal behavior automatically
by extracting communicative cues from both simulated and real
scenarios. Thus, most of these social signal processing frameworks
involve the detection of a set of visual indicators from the
analysis of the participants' body and face. Additionally,
information obtained from speech is commonly
used~\cite{so69562,Vinciarelli20091743,Jayagopi:2010:MGN:2219117.2219993},
as is other information obtained from ambient and wearable sensors
\cite{gatica:multimodal}.

Many of the aforementioned studies demonstrate that indicators of
agreement during communication are highly dependent on social
signals. As such, it is possible to perform an exhaustive analysis
to detect the role played by each participant in terms of
influence, dominance, or submission. For instance, in
\cite{escalera:dominance}, both the interest of observers and the
dominant participants are predicted solely on the basis of
behavioral motion information when looking at face-to-face (also
called \textit{vis-a-vis} or dyadic) interactions. Furthermore,
there are many interdisciplinary, state-of-the-art studies
examining related fields from the point of view of social
computing, some of which are summarized in
\cite{pentland:computation,pentland:honestsignals}.

In most of these frameworks, it can be observed that both ambient
intelligence and egocentric computing methods are defined. Ambient
intelligence refers to electronic environments that are sensitive
and responsive to the presence of people, whereas egocentric
computing refers to the use of wearable devices. Often, existing
techniques of data acquisition make use of interface devices
\cite{Takahashi:interface}, or special items such as gloves
\cite{fang:ttm} to increase recognition accuracy. However, while
these techniques give impressive results in simulated
environments, their use becomes largely infeasible in real-case
scenarios due to their invasiveness and the uncontrollable nature
of the application context. Because of the need to avoid wearing
intrusive egocentric devices, some other ambient sensors that
provide multi-modal data might be considered. In
\cite{gatica:bodycues}, a custom developed system is applied in a
real-case scenario for job interviews. The data acquisition
procedure is performed using different types of camera, by setting
them up in different positions and with different ranges for
capturing visual and depth information. Similarly, scenes with
non-invasive systems have been proposed in other studies, such as
\cite{ponce:bog}, which provides trajectory analyses from body
movements and gestures. Furthermore, audio information has been
analyzed in
\cite{Biel:2011:VCB:2037676.2037690},
with the objective of modeling descriptors for speech recognition.
This can be useful information for measuring the levels of
activity from speech cues, including  detection of
speech/non-speech, interruptions, pauses, or segments obtained
from a speaker diarization process.

The participants involved in these conversational settings usually
appear either sitting or with some parts of their body occluded.
Therefore, from a computer vision point of view, it might be best
just to focus only on the upper body regions. Then, visual feature
extraction techniques can focus on the most significant sources of
information coming from the region of interest, which might be the
face or hands, for example. These regions provide discriminative
behavioral information, or adaptors, which are movements, such as
head scratching, indicative of attitude, anxiety level and
self-confidence \cite{mcneill:hand}; or beat gestures, which are
small baton-like movements of the hands used to emphasize
important parts of speech with respect to the larger discourse
\cite{mcneill:gesture}. However, as explained in
\cite{gatica:bodycues,mehrabian:nonverbal}, body posture is also
found to be an important indicator of a person's emotional state.
Additionally, another potential source of information is provided
by facial expressions
\cite{so69562,Vinciarelli20091743,hernandez:grabcut,RudovicEtAlPAMI12}.

In order to analyze these visual features automatically most
approaches are based on classic computer vision techniques applied
to RGB data. However, extracting discriminative information from
standard image sequences is sometimes unreliable. In this sense,
recent studies have included compact multi-modal devices which
allow $3D$ partial information to be obtained from the scene. In
\cite{shotton:depth}, the authors proposed a system for real-time
human pose recognition including depth information for each image
pixel. In this case, information is obtained by means of a
Kinect\texttrademark\hspace{1mm}device, which estimates a depth
map based on the inverse of time response of an infrared sensor
sampling within the scene. This new source of information, which provides visual $2.5D$ features, has been recently exploited for creating new human pose descriptors by combining different state-of-the-art RGB-depth features ~\cite{hernandez:bovdw}, as well as they are used in a large amount of Human Computer Interaction (HCI) applications \cite{reyesSpherical}.

Once body features are computed, behavioral indicators can be
analyzed by studying their trajectories using pattern recognition
approaches. Some of these methods in this context are based on
dynamic programming techniques such as Dynamic Time Warping (DTW)
\cite{bautista:pdtw} or involve statistical approaches, such as
Hidden Markov Models (HMM) and Conditional Random Fields
(CRF)~\cite{sclaroff:signcrf,Stefan:scalegr,Starner96real-timeamerican}.

Once data from the environment have been acquired and processed to
define a set of behavioral features, they serve as the basis for
modelling a set of communication indicators. For instance, in
\cite{pentland:pfinder}, the authors outline a system for
real-time tracking of the human body with the objective of
interpreting human behavior. However, in the context of
conversations we are particularly interested in behavioral traits
belonging to social signals captured in the communication and in
the interactions between the participants in the VOM sessions. In
this sense, levels of agitation (or energy), activity, stress, and
engagement are analyzed not only from their body movements, but
also from their speech, facial expressions, or gaze directions, in
order to predict behavioral responses.


In this paper, we propose a non-invasive ambient intelligence
framework for the analysis of conversations applied to real-case
scenarios of VOM processes. Due to the ethical constraints imposed
by the legal system, to date no datasets of VOM sessions have been
made publicly available. Here, for the first time, a
novel VOM data set has been collected. In terms of feature
extraction, we extract a set of multi-modal audio-RGB-depth
features and behavioral indicators, which are then used to measure
the degree of receptivity, agreement, and satisfaction using
state-of-the-art machine learning approaches and the ground truth
defined by the mediators in the VOM sessions. As a result, we find
that our technology achieves a high correlation between the most
relevant features obtained by the behavioral indicators and the
information provided by the experts. 


The rest of the paper is organized as follows. Section 2 presents
the material acquired and used in this study. In section 3, the
various system modules are described. Section 4 outlines the
proposal setup and the experimental results. Finally, section 5
concludes the paper.

\section{Data collection} \label{sec:material}

This section describes the data acquisition process, as well as how the annotations of social variables have been defined.

\begin{figure}[t]
    \centering
    \includegraphics[width=7cm,height=7cm]{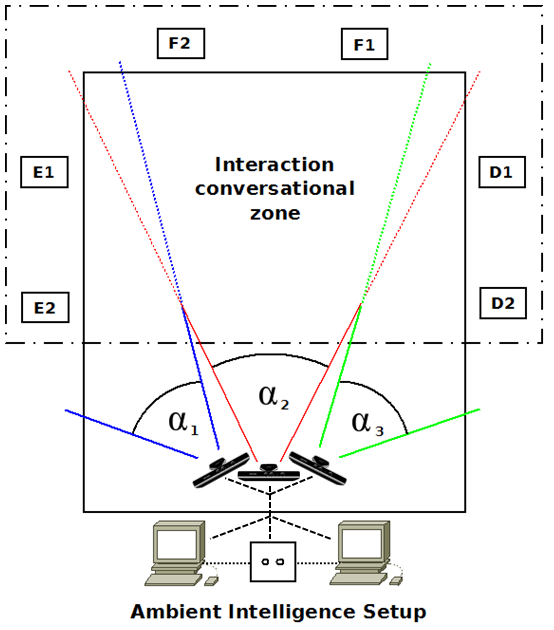}
    \caption{Acquisition architecture. $E1$, $E2$, $F1$, $F2$, $D1$, $D2$ are the participants codified by their respective positions (E: left, F: front, D: right); the angles of view for the different cameras are the same, and hence $\alpha_1 =\alpha_2=\alpha_3$.}
    \label{fig:acq_architecture} 
\end{figure}

First, an environmental study was undertaken in the various rooms
in which recording was to take place, and in which the
non-invasive devices were to be set up. 

Once the environmental study had been completed, decisions
regarding the ethical constraints that had to be satisfied were
taken in order to protect the recorded data. This procedure
involved the drawing up of three fundamental ethical documents:
the researchers' signed undertaking, informed consent, and the
case-codification. 

As the sessions typically involve two or three participants, the
homogeneous distribution of the cameras enabled us to capture at
most two people-per-camera. Specifically, the devices used were
three Kinect\texttrademark\hspace{1mm}sensors and two laptops
(which varied depending on the number of participants). Thus, a
maximum of six people could be recorded\footnote[1]{The
maximum number of people in the recorded sessions was five.}.
Figure \ref{fig:acq_architecture} shows the ambient intelligence
setup with all the elements involved and their distribution.

\begin{figure}[t]
    \centering
    \includegraphics[width=7cm,height=7cm]{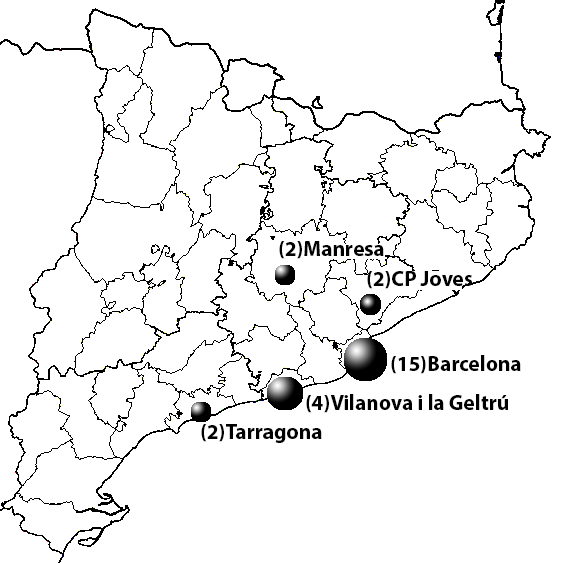}
    \caption{Map with recorded regions marked with dots.}
    \label{fig:map_cat} 
\end{figure}

Recordings were made in various towns and cities of Catalonia.
Figure~\ref{fig:map_cat} shows the distribution of these recording
sites together with the number of sessions recorded in each. Most
recordings were made in the capital city of Barcelona with a total
of 15 sessions, followed by Vilanova i la Geltr\'{u} with a total
of four. Two sessions were recorded in each of Manresa, Tarragona,
and the youth penitentiary center in Granollers. Finally, one
session was recorded in Terrassa.

Thus, 26 VOM sessions, of 35 minutes-per-session, were recorded
(average duration ranging from 20 minutes to 2 hours depending on
the session), in which a mediator engaged in a conversational
process with different parties. Of the total number
of sessions, 15\% were joint encounters, with both parties (victim
and offender) being present in the VOM. The remaining sessions
were individual encounters involving one or other of the parties
and the mediator. As indicated in the VOM program itself, the
participants include both adult men and women, and they fulfill
different participant roles: victim, offender, or mediator. Some
of the sessions also involved accompanying persons, either a
professional from the specific center, or experts in some
particular field relevant to the case under discussion.

Each recorded session\footnote[2]{See an example of the different
modalities and visual extracted visual features in the
\textbf{supplementary video material sample.}} provided
audio-RGB-depth information. These modalities were registered
using the camera parameters, and synchronized between the various
devices through the system clock. The set of images for each
session were recorded at a resolution of 640$\times$480 and at an
average of 12 frames per second (fps), both for RGB and depth
information. Each audio channel, belonging to one of the four
microphones spread out linearly along a multi-array microphone,
processed 16-bit audio at a sampling rate of 16 kHz. The distance
between participants and the Kinect\texttrademark\hspace{1mm}device
was between 1 and 2 meters depending on the specific
characteristics of the recording facility.

As the data protection regulations only allow one
mediator to annotate each session, the annotators were those mediators that had greatest familiarity with the case being dealt with in each session. Only in a few isolated cases there were two mediators in the session.
Thus, in some cases the questionnaires completed by the mediators,
recording their impressions and feelings regarding the party/ies
and the overall sessions, were subsequently confirmed by a second
mediator from the team so as to guarantee the consistency of the
defined ground truth values. The system responses were determined
by considering both the state-of-the-art methods for the study of
behavioral traits in people involved in similar scenarios, as
presented in section~\ref{sec:relatedwork}
~\cite{umbreit:vom,pentland:computation,pentland:honestsignals,Mohammadi:2013:PRP:2522848.2522857,gatica:groupleaders,gatica:bodycues,Aran:2013:OKI:2522848.2522859,so69562,Vinciarelli20091743,Jayagopi:2010:MGN:2219117.2219993,gatica:multimodal,escalera:dominance},
and in the subsequent discussion held with the mediators, taking
into account the aims of their work with the Department of
Justice. Finally, we defined the system's ground truth as:

\begin{itemize}
    \item \textbf{Receptivity}: degree of engagement shown by each party during the session.
    \item \textbf{Agreement}: degree of agreement reached between the parties (quantified globally for each session).
    \item \textbf{Satisfaction}: degree of agreement reached between the parties in relation to the mediator's expectations (quantified globally for each session).
\end{itemize}

These terms do not only depend solely on the individual's
personality, but they are also dependent on the interaction with
other participants, and hence they are a consequence of putting
the behavioral traits into practice. The quantitative nature of
these social responses was validated by a randomly selected
mediator who had not been involved in that case so as to obtain a
more objective evaluation. This approach was likewise applied to
two features describing the evolution in the level of nervousness
manifest by each party at the beginning and at the end of the
process, respectively. Therefore, for each session and for each
party, mediators ranked the observed quantity of these behavioral
indicators from 1 to 5, where 1 is the lowest value and 5 the
highest. Table~\ref{table:data} shows a numerical summary of the
data acquired.

\begin{figure}[b]
    \centering
    \includegraphics[width=8cm,height=7cm]{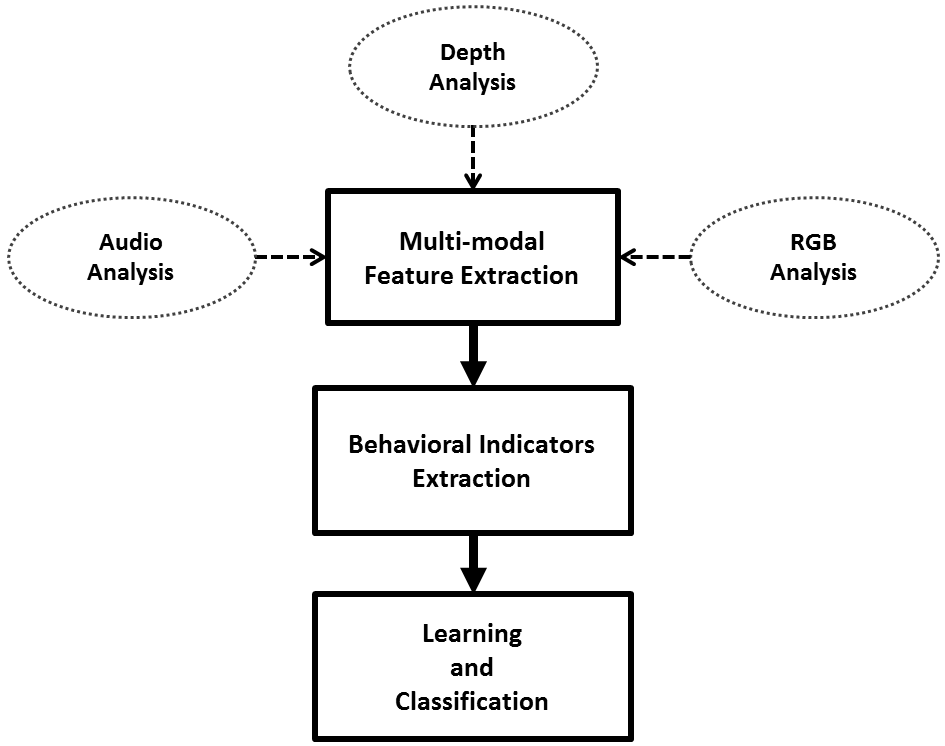}
    \caption{Modules of the proposed system.}
    \label{fig:modules} 
\end{figure}

\begin{table}[h]
    \centering \caption{Summary of data acquired.}
    \begin{tabu}{ |c|r| }
        \hline
        Individual encounters & 22 \\ \hline
        Joint encounters & 4 \\\tabucline[2pt]{-}
        Total sessions & 26 \\ \hline \hline
        Penitentiary centers & 1 \\ \hline
        Office centers & 4 \\\tabucline[2pt]{-}
        Total justice centers & 5 \\ \hline \hline
        Mediators & 7 \\ \hline
        Parties & 30 \\\tabucline[2pt]{-}
        Total n$^o$ participants & 37 \\ \hline \hline
        Total n$^o$ frames & 1,436,400 \\ \hline \hline
        Average n$^o$ minutes/session & 35 \\
        \hline
    \end{tabu}
    \label{table:data} 
\end{table}

\section{Proposed Method}

\begin{figure}[t]
    \centering
    \includegraphics[width=7cm,height=7cm]{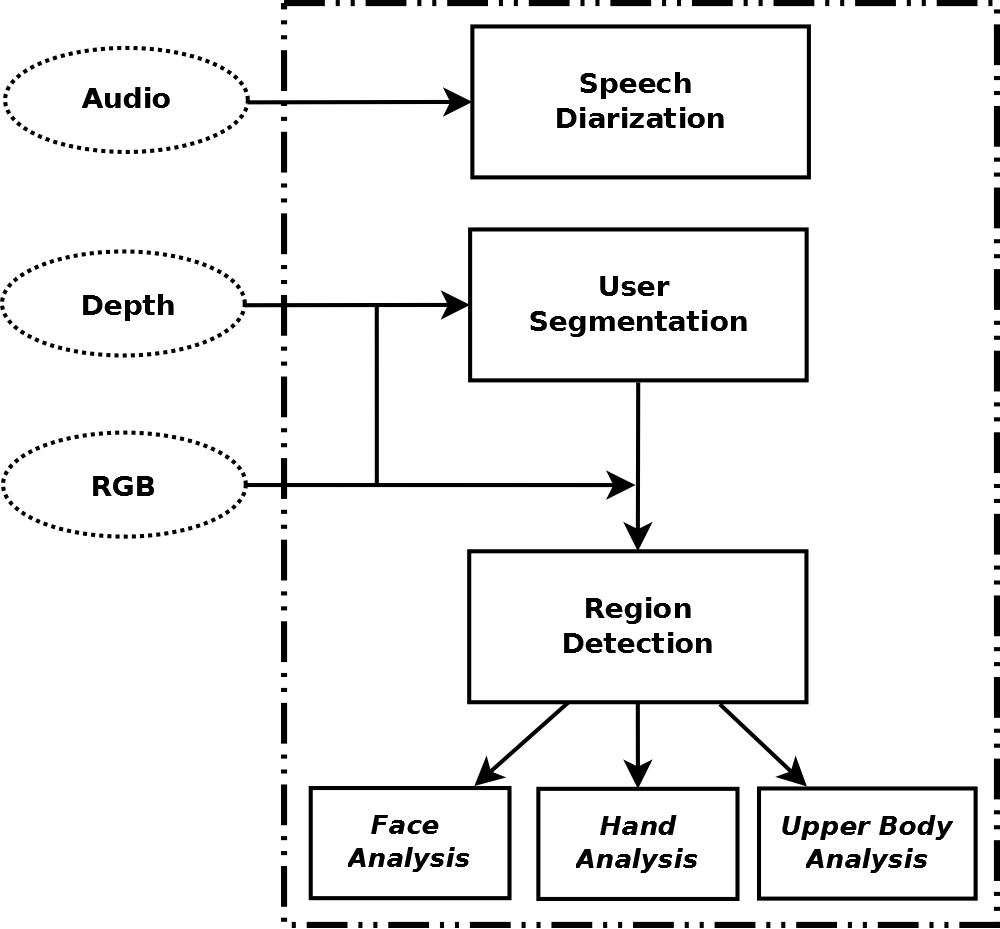}
    \caption{Multi-modal feature extraction module.}
    \label{fig:multimodal} 
\end{figure}

The method proposed for the non-verbal communicative
analysis is described in this section. The framework consists of
three main sequential modules illustrated in
Figure~\ref{fig:modules}. The first module includes the
multi-modal feature extraction from audio-RGB-depth data, which is
described in Figure~\ref{fig:multimodal}. As shown in the scheme,
the steps for obtaining multi-modal features from different
sources of information are the speaker diarization, user
segmentation, and region detection. Once the multi-modal features
have been extracted, they are used to define the behavioral
indicators in the second module, which are used as inputs for the
learning and classification steps in the system's third module.

For all the system's modules, let $V=\{v_1,v_2,...,v_r\}$ be a
set of $r$ recorded VOM sessions and $C=\{c_1,c_2,...,c_s\}$
be a set of $s$ VOM cases. Since a case is divided into one or
more VOM sessions, one session $v \in V$ may belong either to the
same case as another session, or to a different case. Thus, a
subset $\{v_1,v_2\} \in V$ belonging to the first case $c_1$ is
then denoted as $\{v_1^1,v_2^1\}$, where the super-indices
indicate the number of the case.

The rest of this section describing multi-modal feature extraction
is structured as shown in Figure~\ref{fig:multimodal}, where each main block is
described in a different section and where it is dependent on both
the information modality and the previous block. Thus, following
the remaining modules of Figure~\ref{fig:modules}, we describe
the behavioral indicators and finally the learning and
classification of receptivity, agreement, and satisfaction
labels.

\subsection{Audio Analysis}

This section describes the feature extraction applied to the audio
data source. As audio cues are the primary source of information,
they can provide useful evidence for speech production. Although
in noisier environments these cues do not easily distinguish
between the user that is speaking from the other participants (as
a result of overlapping speech), in VOM scenarios this is not such
a problem given the small number of participants. 

\subsubsection{Speaker Diarization} \label{sec:speechDiar}

In order to obtain the audio features, we use a diarization scheme
based on the approach presented in~\cite{Deleglise05thelium}.
These features correspond to state-of-the-art methods for audio
descriptions, which have been successfully applied in several
audio analysis
applications~\cite{Ajmera04robustaudio,Anguera06robustspeaker,Rabiner:1993:FSR:153687}.
The process is described below:

\textbf{Description}: The input audio is analyzed using a
sliding-window of $25$~ms, with an overlap of $10$~ms between
consecutive windows, and each window is processed using a
short-time Discrete Fourier Transform (DFT), mapping all
frequencies to the Mel scale. A more precise approximation of this
scaling for frequencies used in Mel Frequency Cepstral Coefficients
(MFCC) implementations, is represented as:

\begin{equation}
    \hat{f}_{mel}=k_{const}\cdot \log_n \left( 1+\frac{\hat{f}_{lin}}{F_b} \right),
\end{equation}
where $F_b$ and $k_{const}$ are constant values for frequency and scale, respectively. The Koenig scale $\hat{f}_{lin}$ is exactly linear below 1000 Hz and logarithmic above 1000 Hz. In brief, given $N$-point DFT of the discrete input signal $\tilde{x}(n)$,
\begin{equation}
     \tilde{X}(k)=\sum\limits^{N-1}_{n=0}\tilde{x}(n) \cdot \exp \left( \frac{-\tilde{j}2\pi nk}{N} \right), k=0,1,...,N-1,
    \label{DFT}
\end{equation}
a filter bank with several equal height triangular filters is
constructed. Each of these filters has boundary points expressed
in terms of position, which depends on the sampling frequency and
the number of points $N$ in the DFT. Finally, the Discrete Cosine
Transform (DCT) is used to obtain the first $13$ MFCC
coefficients. These coefficients are complemented with the first
and second time-derivatives of the Cepstral coefficients.

\textbf{Speaker segmentation}: Once the audio data are properly
described by means of the aforementioned features, the next step
involves identifying the segments of the audio source which
correspond to each speaker. A first coarse segmentation is
generated according to a Generalized Likelihood Ratio, computed
over two consecutive windows of $2.5$~s. Each block is represented
using a Gaussian distribution, with a full covariance matrix, over
the extracted features. This process produces an over-segmentation
of the audio data into small homogeneous blocks. Then, a
hierarchical clustering is applied to the segments. We use an
agglomerative strategy, where initially each segment is considered
as a cluster, and at each iteration the two most similar clusters
are merged, until the stopping criterion of the Bayesian
Information Criterion (BIC) is met. As in the previous step, each
cluster is modeled by means of a Gaussian distribution with a full
covariance matrix and the centroid distance is used as the link
similarity. Finally, a Viterbi decoding is performed in order to
adjust the segment boundaries. Clusters are modeled by means of a
one-state HMM using GMM as our observation model with diagonal
covariance matrices. Figure~\ref{fig:features} (f) represents an
example of this procedure, showing the clusters where the speech
signal falls at each instant. Since most of the participants
appear in just a single mediation session, we do not learn any
speaker models from the cluster GMMs. Therefore, models extracted
from one session are not used in the diarization process of other
sessions.

\subsection{User Detection} \label{sec:userdetect}

Both RGB and depth data are used for the postural and behavioral
analyses of the parties (for examples of these images see
Figure~\ref{fig:features} (a) and (b), respectively). In this
sense, the first step involves performing a limb-segmentation of
the body based on the Random Forest method
of~\cite{shotton:depth}. This process is performed by computing random offsets of depth features as:
\begin{equation}
    f_\theta(I,\dot{p})=d_I\left(\dot{p}+\frac{\mu}{d_I(\dot{p})}\right)-d_I\left(\dot{p}+\frac{\nu}{d_I(\dot{p})}\right),
    \label{depthFeatures}
\end{equation}
where $d_I(\dot{p})$ is the depth at pixel $\dot{p}$ in image $I$, and parameters $\theta=(\nu,\mu)$ describe offsets $\nu$ and $\mu$. Since offsets are normalized by $\frac{1}{d_I(\dot{p})}$, features are $3D$ translation invariant. Each split node consists of a feature $f_\theta$ and a threshold $\tau$. To classify pixel $\dot{p}$ in image $I$, one starts at the root and repeatedly evaluates Eq. \ref{depthFeatures}, branching left or right according to the comparison to threshold $\tau$.
At the leaf node reached in tree $t$ of the Random Forest, a learned distribution $P_t(l|I,\dot{p})$ over
body part labels $l$ is stored. The distributions are averaged together for all the trees in the forest to obtain the final classification:
\begin{equation}
    P(l|I,\dot{p})=\frac{1}{T}\sum\limits^{T}_{t=1} P_t(l|I,\dot{p}).
    \label{RF}
\end{equation}
Then, we choose body part labels with high probability values so
as to detect those pixels belonging to the person.
Figure~\ref{fig:features} (c) shows a user detection example of applying this segmentation.

Once regions of interest have been located, it is of particular
interest to obtain real-world distance values for certain computed
features so that they are comparable between different subjects.
To do this, we employed a similar procedure to that explained in~\cite{fisher}, which converts the 2D pixels into 3D real-world coordinates using the Kinect\texttrademark\hspace{1mm}depth values. However, since these raw sensor values returned by the depth sensor are not directly proportional to the depth, in ~\cite{fisher}, they scale with the inverse of the depth. Therefore, each pixel $(\dot{x},\dot{y})$ of the depth camera can be projected to metric $3D$ space as:
\begin{equation}
    x = (\dot{x} - \delta_x)\frac{d(\dot{x},\dot{y})}{\kappa_x}, y = (\dot{y} - \delta_y)\frac{d(\dot{x},\dot{y})}{\kappa_y}, z = d(\dot{x},\dot{y}),
    \label{depthtoworld}
\end{equation}
where $(x,y,z)$ will be the real world coordinates, and $\delta_x$, $\delta_y$, $\kappa_x$, $\kappa_y$, the intrinsics of the depth camera. These values will be computed over the detected interest regions in order to define the communicative indicators described in next sections.

\subsection{Region Detection}

This section describes the different feature extraction modules
applied to the visual data source once the user has been
segmented. Specifically, we perform an analysis of the face,
hands, and upper body, as well as visual movements in these
regions during conversations. Moreover, the estimation of head and
body postures provides information about where each person in the
VOM sessions directs their gaze, and the status of the people in
terms of agitation, respectively. Given that it is
not feasible to annotate manually all the regions in several
thousands of frames to measure the recognition rates of the
detection methods used in this section, we propose a heuristic
procedure which maintains the continuity of region detections,
removing false positives and recovering false negative regions,
considerably reducing the manual interaction rate.

\subsubsection{Face Analysis} \label{sec:face}

We are primarily concerned with obtaining the head pose angle of
each of the participants in the session. To do this, we base our
approach on that of \cite{ramanan:aam} which uses a set of face
models. The face model is based on a mixture of trees with a
shared pool of parts $B$, where every facial landmark is modelled
as a part and global mixtures are used to capture topological
changes due to viewpoint. Global mixtures can also be used to
capture gross deformation changes for a single viewpoint, such as
changes in expression.
Each tree is written as $T_m=(B_m,E_m)$ as a linearly-parameterized, tree-structured pictorial structure \cite{YangR_CVPR_2011}, where $m$ indicates a mixture and $B_m \subseteq B$. Let $I$ be an image, and $\rho_i=(x_i,y_i)$ the pixel location of part $i$. Configuration of parts $L=\{\rho_i:i\in B\}$ are then scored as:

\begin{equation}
    S(I,L,m)=App_m(I,L)+Shape_m(L)+\alpha^m,
    \label{Smodel}
\end{equation}
\begin{equation}
    App_m(I,L)=\sum\limits_{i\in B_m} w_i^m\cdot \phi(I,\rho_i),
    \label{App}
\end{equation}
\begin{equation}
    Shape_m(L)=\sum\limits_{ij\in E_m} \gamma^m_{ij} dx^2 + \eta^m_{ij} dx + \varrho^m_{ij} dy^2 + \varphi^m_{ij} dy.
    \label{Shp}
\end{equation}
Eq.~\ref{App} sums the appearance evidence for placing a template $w^m_i$ for part $i$, tuned for mixture $m$, at location $\rho_i$. The feature vector $\phi(I,\rho_i)$ (e.g., HoG descriptor) is extracted from pixel location $\rho_i$ in image $I$. Eq.~\ref{Shp} scores the mixture-specific spatial arrangement of parts $L$, where $dx = x_i-x_j$ and $dy=y_i-y_j$ are the displacement of the $i$th part relative to the $j$th part. Each term in the sum can be interpreted as a spring that introduces spatial constraints between a pair of parts, where the parameters $(\gamma,\eta,\varrho,\varphi)$ specify the rest location and rigidity of each spring. Finally, the last term $\alpha^m$ is a scalar bias or "prior" associated with viewpoint mixture $m$. Eq.~\ref{Smodel} requires a separate template $w^m_i$ for each mixture/viewpoint $m$ of part $i$. However, parts may look consistent across some changes in viewpoint. In the extreme cases, a "fully shared" model would use a single template for a particular part across all viewpoints, $w^m_i=w_i$. A continuum between these two extremes is explored, written as $w^{f(m)}_i$, where $f(m)$ is a function that maps a mixture index (from 1 to $M$) to a smaller template index (from 1 to $M'$). Various values of $M'$ are explored: no sharing $(M'=M)$, sharing across neighboring views, and sharing across all views $(M'=1)$.

The detection of the head pose angle is performed by averaging HOG
feature as a polar histogram over 18 gradient orientation
channels, as computed from the entire PASCAL 2010 data set
\cite{Everingham:2010:PVO:1747084.1747104}. In
Figure~\ref{fig:features} (d) we can visualize the set of computed
features plotted on the detected face. 

\begin{figure}[t]
    \centering
    \includegraphics[width=8cm,height=15cm]{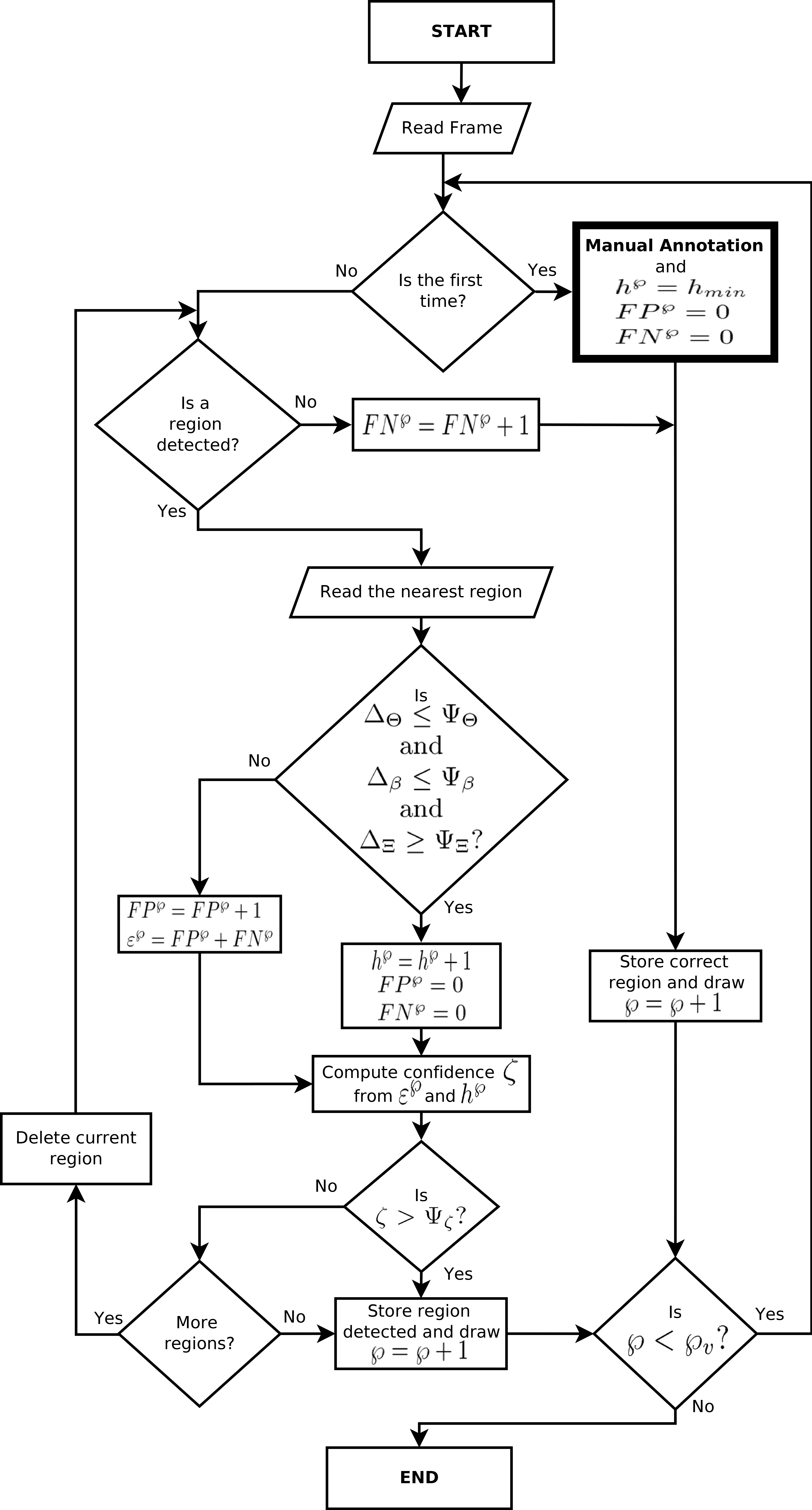}
    \caption{Flowchart of the heuristic procedure applied to each frame. The total number of people that appear in the current video $v$ is denoted by $\wp_v$. Constraints of the main condition at the center of the flowchart are denoted by $\Delta_\Theta$, $\Delta_\beta$, $\Delta_\Xi$, and their respective thresholds $\Psi_\Theta$, $\Psi_\beta$, $\Psi_\Xi$. The counting variables are $FN^\wp$, $FP^\wp$, $h^\wp$, representing the accumulated number of false negatives, false positives, and hits for the current person $\wp$. They are used to compute the confidence $\zeta$ from the accumulated detection errors $\varepsilon^\wp$ and the hits $h^\wp$, and to decide whether the current detected region has to be stored or discarded through the threshold $\Psi_\zeta$.}
    \label{fig:heuristic_chart} 
\end{figure}

\begin{figure*}[t]
    \centering
    \includegraphics[width=\textwidth]{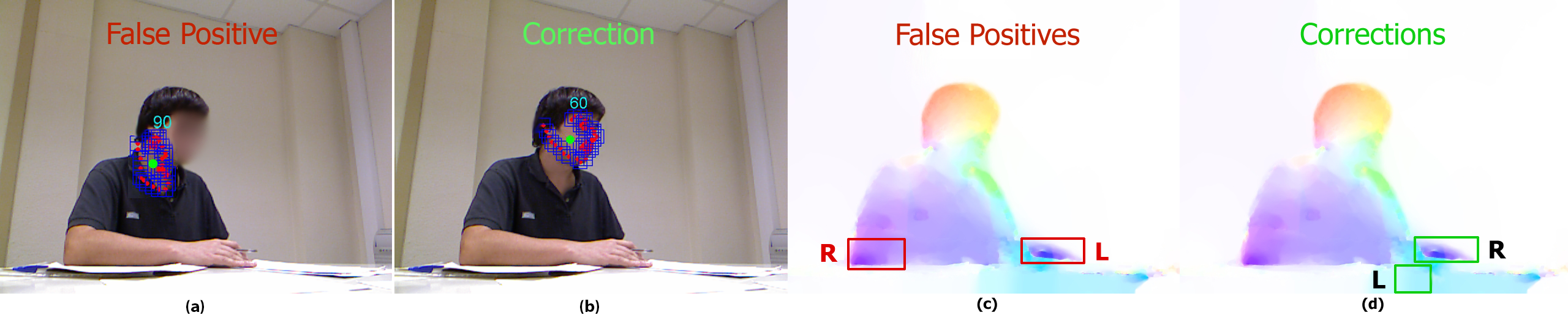}
    \caption{Examples of how the semi-automatic heuristic procedure works on two pairs of frames of a session. The correction of false positives and false negatives is shown, improving the continuity of the detection of positive regions of interest between consecutive frames. Image (a) shows a false positive detection for the face region, whereas image (b) shows its correction with the proper fitting. Image (c) shows false positive detections for the hand regions, choosing those blobs obtained by means of skin segmentation having highest optical flow with respect to the previous frame. Image (d) shows the correction of these regions by comparing them with the positive hand detections from the previous frame.}
    \label{fig:heur_results} 
\end{figure*}

While face detection takes place for each tested image, a
semi-automatic heuristic procedure is proposed in order to improve
the continuity of positive detections of regions of interest in
the person between consecutive frames, and to correct possible
erroneous detections due to the inherent difficulties of the
problem at hand. Figure~\ref{fig:heuristic_chart} shows the
flowchart of the procedure applied to each frame. In short, it
consists of a temporal filtering methodology of detected regions
(faces) between one-by-one consecutive frames. It is based on
three main constraints that enable us to choose the detected
regions in the current frame by comparison to the previous one:
offset pixels produced by the mass centers, offset angle produced
by head poses, and the size difference factor produced among the
region areas. Thus, three thresholds $\Psi_\Theta$, $\Psi_\beta$,
and $\Psi_\Xi$ are respectively used to discriminate the occurred
cases on each constraint, whose values may vary depending on the
session conditions. Moreover, there are three counting variables
that accumulate, for each person, the number of correct detections
(hits) $h^\wp$, false positives $FP^\wp$ and false negatives
$FN^\wp$. Then, a confidence $\zeta$ is computed from $h^\wp$ and
the sum of false detections $\varepsilon^\wp$ to decide whether
the current detected region has to be stored or discarded by means
of the threshold $\Psi_\zeta$. These counting variables are highly
dependent on constraint thresholds, as they make the system more
or less restrictive when choosing detected regions. Therefore, a
trade-off between constraint thresholds and control thresholds
should be reached when assigning their values in order to assure
the continuity of positive region of interest detections for that
person (even though the method could not detect any region in the
image), and to decide whether a manual annotation is required to
re-initialize the detection process in the (approximately) desired
frequency rate. Images (a) and (b) of
Figure~\ref{fig:heur_results} show an example of how the heuristic
procedure corrects a false positive detection on the face region
for a given frame. In the end, so that all image sequences were
annotated with the head pose analysis, we found that the use of
the proposed heuristic reduced manual interactions by a ratio of
up to five.

\subsubsection{Hand Analysis} \label{sec:hand}

Given that the skeletal model computed from the person
segmentation image \cite{shotton:depth} does not offer an accurate
fit of the hand joints in our particular scenario, we designed a
semi-automatic procedure for hand detection.

First, hands are manually annotated in the starting frames of each
session to perform posterior color segmentation for the rest of
the frames. In this way, a GMM is learned with the marked set of
most significant pixels, defining the skin color model of the
person. Then, subsequent frames are
tested within the GMM built using a threshold $\vartheta$,
discriminating those pixels belonging to the skin color from those
belonging to the background. The resulting blobs are filtered
using mathematical morphology closing operation with a $3\times 3$
square structured element to discard noise and to obtain smoother
regions. Once the set of blobs has been obtained, we need to
choose those two candidates that belong to the hand regions. This
is performed by computing the optical flow between
consecutive frames, which allows to discard noise in those
cases in which we obtain more than two blobs by retaining those
with higher movement. Figure~\ref{fig:features} (e) illustrates
this procedure.

On the other hand, we use the same heuristic procedure as that
 applied to the face analysis step for choosing the two best hand
 candidates. Images (c) and
(d) of Figure~\ref{fig:heur_results} show an example of how the
heuristic procedure corrects false positive detections on the
regions of the hands. The incorrect regions detected in the first
instance are the blobs presenting the highest optical flow, and
then the heuristic procedure corrects these regions by comparing
them with the hand regions obtained from the previous frame. As in
face detection, manual annotation may be required in those cases
where the heuristic procedure needs to be re-initialized. For this
task, an interface has been designed for the manual annotation of
the hand regions for the set of frames in which this occurs. When
the user makes any annotation, the GMM color model is newly
re-constructed at this frame using the marked pixel positions, and
the whole process is repeated. In this case, using the proposed
heuristic we also found similar reduction regarding manual
interaction effort as in the case of face region detection.

Once we have obtained the blobs belonging to the hand regions, the
extremes with higher optical flow magnitude are used to obtain
$2D$ hand positions. Finally, these positions are transformed to
$3D$ real world coordinates using the conversion of Eq.~\ref{depthtoworld}. 

\subsubsection{Upper Body Analysis} \label{sec:body}

As presented above, the probability of each pixel of an image
belonging to a labeled body part is computed 
in Eq.~\ref{RF} using depth features of Eq.~\ref{depthFeatures}.
This information is used for the subsequent calculation of optical flow on RGB images
where the upper body region appears. Therefore, each pixel
$\dot{p}$ of the image $I$ detected by Random Forest with high
probability of being part $l$ of the person is used to calculate
the optical flow. Finally, an average $\bar{\sigma}$ of optical
flow is computed for the upper body region. An example of user
detection where upper body region is highlighted is shown in
Figure~\ref{fig:features} (c). Average optical flow $\bar{\sigma}$
is later used to define behavioral indicators.

\subsection{Behavioral Indicators} \label{sec:behavindic}

Once the multi-modal features have been extracted, we
can use them to build a set of behavioral indicators that reveal
communicative cues about each party involved in the VOM process.
This information is of great interest in detecting the response of
subjects to certain feelings or emotional states during the
conversation \cite{knapp1997nonverbal}. In
particular, since the behavioral cues of the mediators are not of
interest for our purposes here, we focus mainly on those of the
parties.

Thus, in this section, we describe the set of behavioral
indicators constructed from the output of the different blocks
shown in Figure~\ref{fig:multimodal}, which define the final
feature vector for each party within the VOM process.

\subsubsection{Target Gaze Codification}

The head pose and the face is obtained by applying
the methodology explained in section \ref{sec:face}. Therefore, in
a given session, we compute the correlations between the head pose
angles belonging to each participant and the positions taken by
the rest of the participants in that session. This procedure is
performed in order to identify the visual focus of attention among
the different participants in the conversation
\cite{Aran:2013:OKI:2522848.2522859,IJCV:Looking,SMCB:4625982}. For this purpose, different
ranges are assigned to each participant in terms of angle limits.
Given that the participants belonging to the same party are seated
in adjacent positions (see acquisition architecture in
Figure~\ref{fig:acq_architecture}), each range represents a
possible participant vision field of his/her gaze towards the
target party. Thus, given a frame of the session and a
participant, if his/her head pose angle falls within a particular
range, then the party found within that range is identified as the
target gaze of this participant for that frame, which means the
participant is looking at this party. Since sessions have
different setups, they may consist of one or two parties (and the
mediator), each with a different number of participants.
Therefore, the ranges require manual assignment depending on each
session setup. Then, the target gazes are automatically identified
for all the frames of the session. Table~\ref{table:gazes}
describes all the target gaze combinations that can be found in
any possible session setup. Note that we codify the target gazes
with values $1$ or $0$ depending on whether a person is looking
at someone or not, respectively. The table focuses on the target gazes among the parties rather than isolated participants. For the mediator and party columns, the first and third subcolumn correspond to gazes towards the parties and the second to gazes towards the mediator. Since we focus the attention on the behavior of the parties, if there is a session with more than one mediator, the gazes among the mediators are discarded, and hence they're codified as $0$. Thus, the value $1$ may be assigned to the different parties in the session, described as:

 \begin{itemize}
    \item P: Looking at a party, whether it is the offender (Off) or the victim (Vic). If there is more than one party (case of a joint encounter), P changes on the mediator column by 'Off' or 'Vic' depending on the party.
    \item M: This party is looking at the mediator.
    \item MP: This party is looking at the same party.
    \item MO: This party is looking at the other party.
 \end{itemize}

\begin{table}[h]\scriptsize
    \centering \caption{All possible combinations for the codification of target gazes. Each party }
    \begin{tabular}{|l|c|c|}
      \hline
       & \textbf{M: Mediators} & \textbf{P: Party} \\ \hline
      1 party P with only 1 person & P $|$ 0 $|$ 0 & 0 $|$ M $|$ 0 \\ \hline
      1 party P with several people & P $|$ 0 $|$ 0 & MP $|$ M $|$ 0 \\ \hline
      2 parties with only 1 person in the party P & Off $|$ 0 $|$ Vic & 0 $|$ M $|$ MO \\ \hline
      2 parties with several people in the party P & Off $|$ 0 $|$ Vic & MP $|$ M $|$ MO \\
      \hline
    \end{tabular}
    \label{table:gazes}
\end{table}

As an example to clarify the table~\ref{table:gazes}, consider the third row case (2 parties with only 1 person in the party P) where you are a mediator, and you have another colleage who is also a mediator. Thus, mediators might be looking either at the offender (Off) party or at the Victim (Vic) party, since the gazes among the mediators are not considered (middle subcolumn of the mediators are always $0$). On the other hand, the party P that has only 1 person might be looking either at the mediators (M) or at the other party (MO).

\begin{figure*}[t]
    \centering
    \includegraphics[width=\textwidth]{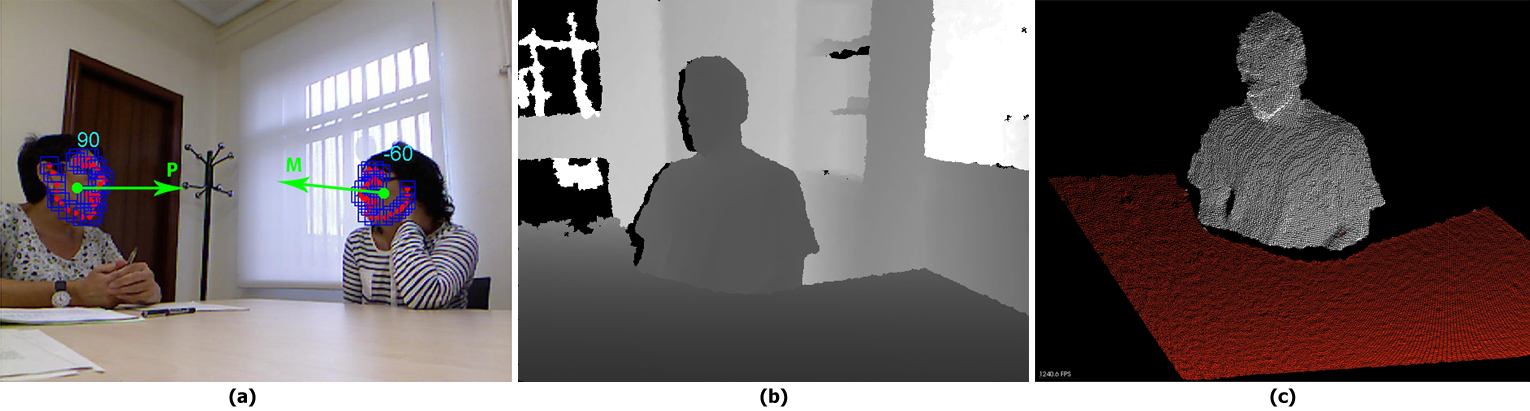}
    \caption{Visual instances of some situations where behavioral indicators are detected in VOM sessions. Image (a) shows the detection of crossed gazes between the mediator and the other participant. Images (b) and (c) show a depth image and its segmentation for the person (white point cloud) and the table (red point cloud), respectively, which is used to detect a situation in which the target subject appears with his or her hands under the table.}
    \label{fig:indicators} 
\end{figure*}

Figure~\ref{fig:indicators} (a) shows an example of crossed gazes
between the mediator and a party in a real VOM session. Finally,
we use these binary values to compute the time percentages of
target gazes for each party. Therefore, for any given party, there
exists a total of $4$ indicators for representing all possible
combinations of target gazes.

\subsubsection{Agitation Estimation}

As explained in section~\ref{sec:hand}, $3D$ positions belonging
to the hand regions are computed from the extreme positions of
higher optical flow. From these positions, we are able to quantify
the movement for each region between consecutive frames. For this
purpose, let $F=\{\iota_1,\iota_2,\iota_3,...,\iota_\lambda\}$ be
a set of consecutive frames. This set of frames belongs to a video
session $v\in V$, being $\lambda=r$ the maximum length of the set.
Then, for each region we compute the average agitation over all
the frames $\iota\in F$ as:
\begin{equation}
    A_h=\frac{1}{\lambda}\sum\limits^\lambda_{\iota=1} \Delta_h^\iota,
\end{equation}
where $\Delta_h^\iota=\Delta_p^\iota+\Delta_q^\iota$
are the displacements among $3D$ positions of hands $h$ (left $p$ and
right $q$) between frames $\iota$ and $\iota-1$, computed using
Euclidean distance. Therefore, $A_h$ contains the accumulated
average of displacements produced by both hands between frames
$F$.

On the other hand, in section~\ref{sec:body} we explained how the
average optical flow $\bar{\sigma}$ is obtained for the upper body
region. Therefore, if we denote as $\bar{\sigma}_\iota$ the
average optical flow of the upper body for a given frame $\iota\in
F$, then:
\begin{equation}
    A_b=\frac{1}{\lambda}\sum\limits^\lambda_{\iota=1} \bar{\sigma}_\iota,
\end{equation}
where $A_b$ contains the accumulated average of optical flow
produced by the upper body between frames $F$.

In short, for each party and session, agitation
averages are computed over processed frames, with a total of $8$
agitation indicators, both alone and combined with the other
indicators previously calculated. The idea of combining these
indicators with other behavioral features is inspired by
\cite{lookingSpeaking,escalera:dominance}. In this case, we
consider a combination between the features describing the
agitation from the upper body and those describing the hands while
looking at the participants (see list of Table~\ref{table:indiclist}).

\subsubsection{Posture Identification}

From the $3D$ body position, we detect the body posture as one behavioral indicator, which may describe the engagement (or
involvement) of the party within the VOM session. Our description
of body posture is classified into three main positions (tilted
backward, normal, tilted forward), where the posture selected is
the one that has the most occurrences over the processed frames.

In addition, $3D$ hand positions are used to detect where the
hands are along the processed frames, in terms of average and time
percentages. In particular, we discriminate three cases (i.e., $3$
indicators): hands together, hands touch the face, and hands under
the table. This is done in a similar way as for the agitation
estimation, using Euclidean distance computed over $3D$ positions.

\begin{itemize}
    \item Hands together: We compute for each frame the distance between left and right hand positions belonging to the target subject, and we consider the frames where the distance values are below that of the threshold $\Psi_\Theta$. Finally, we compute the time percentage for those frames where the target subject appears with their hands together.
    \item Hands touch the face: We compute for each frame the distance between each hand position and the position belonging to the face center of mass obtained in section~\ref{sec:face}. Then, we consider the frames where the distance values are below that of the threshold $\Psi_\Theta$. Finally, we compute the average distance for those frames where the target subject appears with their hands touching their face.
    \item Hands under the table: For each frame, we first perform a segmentation of the tables using~\cite{Rusu_ICRA2011_PCL} to obtain planar objects within images. Then, we compare the $3D$ positions of both hands with the position of the tables in order to discriminate the two possibilities where the hands may appear under or above the table. Finally, we compute the time percentage for those frames where the target subject appears with their hands under the table. Figure~\ref{fig:indicators} (b) and (c) illustrate an example of this procedure, showing respectively the input depth image and its segmentation.
\end{itemize}

\subsubsection{Speech Turns/Interruptions Detection}

The speaker diarization process of section~\ref{sec:speechDiar}
detects time segments belonging to each participant in the VOM
process. In order to extract the degree of interaction, we not
only use the length of time during which each participant speaks,
but we also count the number of turns in each session. This
enables to differentiate between a session where each party
expresses its position from a session in which a conversation is
maintained between the VOM participants. Apart from the
quantification of turn taking, a relevant indication in the social
communication analysis is the detection of interruptions, which
are related to the dominance and respect between two
persons~\cite{s120201702}. Using the time between turns, we
compute the percentage of turns in which a participant interrupts
another one. For instance, in the first three turns of
Figure~\ref{fig:features} (f) a participant (red) interrupts the
mediator (green), while the mediator waits until the other
participant ends his turn before starting to speaking again.

\subsection{Classification} \label{classif}

The total number of behavioral indicators is 36,
which define the feature vector for each sample in our data set.
Here, we define a sample as each party participating in a VOM
session $v\in V$. Thus, if a
session $v$ involves two parties and the mediator, we introduce
one sample of 36 features for each of the two parties. On the
other hand, if a session involves just one party and the mediator,
we introduce only one sample corresponding to the party involved.
Table~\ref{table:indiclist} summarizes the behavioral indicators
and gives a brief description.

\begin{table}[h]
    \centering \caption{Summary of behavioral indicators defining each feature vector. The last two features $f_{35}$ and $f_{36}$ are obtained from survey answers provided by mediators.}
    \begin{tabular}{|c|l|}
      \hline
      \textbf{Feature} & \textbf{Brief description} \\ \hline
      $f_1$ & Party's role within the VOM session (victim or offender)\\ \hline
      $f_2$ & This party looks at the other \\ \hline
      $f_3$ & The other party looks at this party \\ \hline
      $f_4$ & This party looks at the mediator \\ \hline
      $f_5$ & The mediator looks at this party \\ \hline
      $f_6$ & Body posture inclination of this party \\ \hline
      $f_7$ & Gender of the mediator \\ \hline
      $f_8$ & Gender of this party  \\ \hline
      $f_9$ & Gender of the other party \\ \hline
      $f_{10}$ & Age of the mediator \\ \hline
      $f_{11}$ & Age of this party \\ \hline
      $f_{12}$ & Age of the other party  \\ \hline
      $f_{13}$ & Session type (individual/joint encounter) \\ \hline
      $f_{14}$ & Upper body agitation of this party \\ \hline
      $f_{15}$ & Upper body agitation of this party while looking at the other party \\ \hline
      $f_{16}$ & Upper body agitation of this party while looking at the mediator \\ \hline
      $f_{17}$ & Hands agitation of this party \\ \hline
      $f_{18}$ & Hands agitation of this party while looking at the other party\\ \hline
      $f_{19}$ & Hands agitation of this party while looking at the mediator \\ \hline
      $f_{20}$ & Hands agitation of the mediator while looking at this party \\ \hline
      $f_{21}$ & Hands agitation of the other party while looking at this party \\ \hline
      $f_{22}$ & This party has the hands together \\ \hline
      $f_{23}$ & Hands of this party touches his/her face \\ \hline
      $f_{24}$ & This party has the hands under the table \\ \hline
      $f_{25}$ & Mediator speaking time  \\ \hline
      $f_{26}$ & Speaking time of this party \\ \hline
      $f_{27}$ & Speaking time of the other party \\ \hline
      $f_{28}$ & Mediator speaking turns \\ \hline
      $f_{29}$ & Speaking turns of this party \\ \hline
      $f_{30}$ & Speaking turns of the other party\\ \hline
      $f_{31}$ & Mediator interrupts this party \\ \hline
      $f_{32}$ & This party interrupts the mediator \\ \hline
      $f_{33}$ & This party interrupts the other party \\ \hline
      $f_{34}$ & The other party interrupts this party \\ \hline
      $f_{35}$ & Nervousness of this party at the beginning \\ \hline
      $f_{36}$ & Nervousness of this party at the end \\
      \hline
    \end{tabular}
    \label{table:indiclist} 
\end{table}

As explained in section~\ref{sec:material}, the observations of
the classification task are the accuracies achieved by the system
when predicting receptivity, agreement, and satisfaction. Then,
the correlation can be observed between the observations predicted
by the system and the impressions recorded by the mediators. These
opinions are quantified values of receptivity, agreement, and
satisfaction presented in relation to the parties involved in the
VOM session, and represent the ground truth of our system. The
ground truth values are assigned to each sample of the data set.
Since agreement and satisfaction are globally assigned for each
session, those sessions containing two parties will share the same
ground truth labels of agreement and satisfaction for both
generated samples, meanwhile the receptivity ground truth value is
assigned to each sample (party) independently.

Learning is then performed on these samples and their
features as a binary classification problem, grouping into two
classes the quantifications performed by the mediators. To do
this, we employ four classical techniques from the machine
learning field: AdaBoost \cite{Freund96experimentswith}, Support
Vector Machines (SVM) using a Radial Basis Function
(RBF) \cite{CC01a:libsvm}, Linear Discriminant Analysis (LDA),
and three kinds of Artificial Neural Networks (ANN), in particular
Probabilistic Neural Networks (PNN)~\cite{23887}, and
Cascade-Forward (CF) and Feed-Forward neural networks
(FF)~\cite{CFFFNN}. In addition to the binary classification
analysis we also conduct a regression study using epsilon-SVR
(Support Vector Regression)~\cite{CC01a:libsvm} in order to
predict continuous quantifications of the three labels.

\section{Experiments}

This section describes the experiments performed when using the
behavioral indicators summarized in Table~\ref{table:indiclist}.
First, we describe the setting and validation measurements, before
outlining the experiments performed.

\subsection{Setting and validation measurements} \label{sec:settings}

The parameters used in the heuristic procedure were experimentally
set to ranges $\Psi_\Theta\in[50,120]$, $\Psi_\beta\in[30,60]$,
and $\Psi_\Xi\in[0.1,0.3]$, depending on the session, and the
standard value $\Psi_\zeta=0.5$ for all sessions.

The measurements for the features
$\{f_2,f_3,f_4,f_5,f_{22},f_{24},f_{25},f_{26},f_{27}\}$ are time
percentages, whereas the features
$\{f_{14},f_{15},f_{16},f_{17},f_{18},f_{19},f_{20},f_{21},f_{23}\}$
contain averaged values of optical flow or distances, both taking
into account the processed frames of a session. Features
$\{f_{28},f_{29},f_{30},f_{31},f_{32},f_{33},f_{34}\}$ are turn
taking percentages, where a turn means that the speaker changes.

Finally, the remaining features are codified into discrete values,
either quantified to ranges $[1,2]$ or $[1,3]$ integer values for
features $\{f_1,f_6,f_7,f_8,f_9,f_{10},f_{11},f_{12},f_{13}\}$, or
to range $[1,5]$ integer values for the nervousness features
$\{f_{35},f_{36}\}$. Each party of a video session $v$ is a sample
for the classification task, and the total number of used samples
is $28$.

In addition, an alternative was implemented where some features
are divided into two -one belonging to the first half, the other
to the second half of the session-. This procedure was initially
performed to identify behavioral changes in subjects during the
different halves of the session. However, no significant
differences were found and, hence, we finally used the set of
features without any temporal segmentation.

Learning is performed using leave-one-out
validation, keeping one sample out of the testing each time. Since
the total number of samples is small and the ground truth values
are quantified within ranges $[1,5]$ (as for the nervousness
features), we simplified the problem by grouping the different
response degrees into binary groups, but we also performed a
posterior regression analysis. In the case of a binary setup, the
value $3$ can be considered as being either high or low. For this
reason, we ran the experiments twice to test each grouping case, as we show in the result Tables~\cref{table:resallfeat1,table:resallfeat2,table:resoutnerv1,table:resoutnerv2}:

\begin{itemize}
\item First grouping case: Degrees of quantification $\{1,2,3\}$ versus $\{4,5\}$.
\item Second grouping case: Degrees of quantification $\{1,2\}$ versus $\{3,4,5\}$.
\end{itemize}

In our experiments, we awarded the standard value of
$50$ to the number of decision stumps in the AdaBoost technique.
For the SVM-RBF and epsilon-SVR, we experimentally set the cost,
gamma, and epsilon parameters by means of the leave-one-out
validation for each social response and minimizing regression
deviation on the training set. Finally, we applied the same tuning
procedure for the three standard neural network parameters:
a Probabilistic Neural Network (PNN) with a spread value of $0.1$
for the radial basis functions, and Cascade-Forward (CF) and
Feed-Forward (FF) neural networks, both with a single hidden layer
with $10$ neurons values and Levenberg-Marquardt back-propagation
training function. The results obtained are shown in terms of
accuracy percentages.

\subsection{Results and discussion}

Due to the sensitive nature of the VOM process, never before (to
the best of our knowledge) have mediators recorded their sessions
so that they might subsequently analyze the cases. In this
respect, therefore, the first results to emerge from this study
are the session videos themselves, which are valuable materials
via which the mediators can share their experiences and obtain
feedback to improve their mediation skills.

\begin{table}[t]
    \centering \caption{Accuracy considering the first grouping case and all features.}
    \begin{tabu}{|c|r|r|r|r|r|r|r|}
      \hline
      \rowfont[c]{} Label & AdaBoost & LDA & PNN & CF & FF & SVM \\ \hline
      Satisfaction & 57\% & 32\% & 57\% & 57\% & \textbf{86\%} & 57\% \\ \hline
      Agreement & 50\% & 54\% & 64\% & 64\% & 75\% & 64\% \\ \hline
      Receptivity & 64\% & 50\% & 71\% & 71\% & 68\% & 75\% \\
      \hline
    \end{tabu}
    \label{table:resallfeat1} 
\end{table}

\begin{table}[t]
    \centering \caption{Accuracy considering the second grouping case and all features.}
   \begin{tabu}{|c|r|r|r|r|r|r|r|}
      \hline
      \rowfont[c]{} Label & AdaBoost & LDA & PNN & CF & FF & SVM \\ \hline
      Satisfaction & 82\% & 43\% & 21\% & 82\% & 75\% & 82\% \\ \hline
      Agreement & 71\% & 43\% & 29\% & 71\% & 75\% & 75\% \\ \hline
      Receptivity & 75\% & 36\% & 39\% & 68\% & 75\% & 61\% \\
      \hline
    \end{tabu}
    \label{table:resallfeat2} 
\end{table}

As described in section~\ref{sec:face}, user manual interactions
are a key requirement in our proposed semi-automatic system for
improving the continuity of detections of positive regions of
interest between consecutive frames. For our sessions, the average
frequency rate of manual annotations required is $1$ for each
$2000$ frames based on the above parameters. This means that when
using these parameters the feature extraction procedures for both
hands and faces offer high performance.

\begin{table}[t]
    \centering \caption{Accuracy considering the first grouping case and withholding the nervousness features.}
    \begin{tabu}{|c|r|r|r|r|r|r|r|}
      \hline
      \rowfont[c]{} Label & AdaBoost & LDA & PNN & CF & FF & SVM \\ \hline
      Satisfaction & 57\% & 57\% & 57\% & 68\% & 64\% & 57\% \\ \hline
      Agreement & 50\% & 43\% & 64\% & 57\% & 71\% & 64\% \\ \hline
      Receptivity & 68\% & 46\% & 71\% & 75\% & 68\% & 75\% \\
      \hline
    \end{tabu}
    \label{table:resoutnerv1} 
\end{table}

\begin{table}[t]
    \centering \caption{Accuracy considering the second grouping case and withholding nervousness features.}
    \begin{tabu}{|c|r|r|r|r|r|r|r|}
      \hline
      \rowfont[c]{} Label & AdaBoost & LDA & PNN & CF & FF & SVM \\ \hline
      Satisfaction & 82\% & 61\% & 21\% & 71\% & \textbf{86\%} & 82\% \\ \hline
      Agreement & 71\% & 57\% & 29\% & 71\% & \textbf{79\%} & 75\% \\ \hline
      Receptivity & \textbf{79\%} & 46\% & 39\% & 64\% & 71\% & 61\% \\
      \hline
    \end{tabu}
    \label{table:resoutnerv2} 
\end{table}

The predictions addressed in our classification task focus on
three indicators: the degree of receptivity of the parties, the
level of agreement reached, and the degree of mediator
satisfaction. Tables~\ref{table:resallfeat1} and
\ref{table:resallfeat2} show the results obtained when employing
the different techniques and using the complete set of behavioral
indicators in Table~\ref{table:indiclist}. 
The most accurate results for the three responses are shown in bold, showing both which classifier and which grouping case give the best performance for each feature description. Note that as the features of nervousness are subjective indicators that are not
automatically computed, we repeated the experiments without these
two features $\{f_{35},f_{36}\}$. These results are shown in
Tables~\ref{table:resoutnerv1} and \ref{table:resoutnerv2}, where
the prediction is also analyzed under the grouping hypotheses.
Once again, the results show a correlation between the features
extracted and the categories selected. The percentage degree of
accuracy in the predictions is then compared for the different
techniques: AdaBoost, SVM, LDA, PNN, CF, and FF. It can be noted
that, except for PNN and LDA (which are not good techniques for
use with our dataset), all the classifiers are able to make
predictions about the random decision. This indicates that there
is a correlation between the captured data and the information
that we want to predict. The most predictable social
response is that of satisfaction, presenting an accuracy of 86\%
with the FF, followed by 82\% with AdaBoost, SVM, and CF. The best
result when predicting agreement was an accuracy of 79\% with FF
and, similarly, when predicting receptivity, the best accuracy was
79\% with AdaBoost. These results are quite significant since most
of the sessions presenting high values for this combination of
responses resulted in satisfactory VOM outcomes. However, since
the number of samples is, in general, small, all responses vary in
their performance depending on the grouping hypothesis, despite
the low level of presence of the $3$-value among the quantitative
responses. This means that the uncertainty of the mediator when
assigning a value of $3$ to the answers tends to add noise to the
overall data with respect to the evaluation.

\begin{figure}[ht]
    \centering
    \includegraphics[width=8cm]{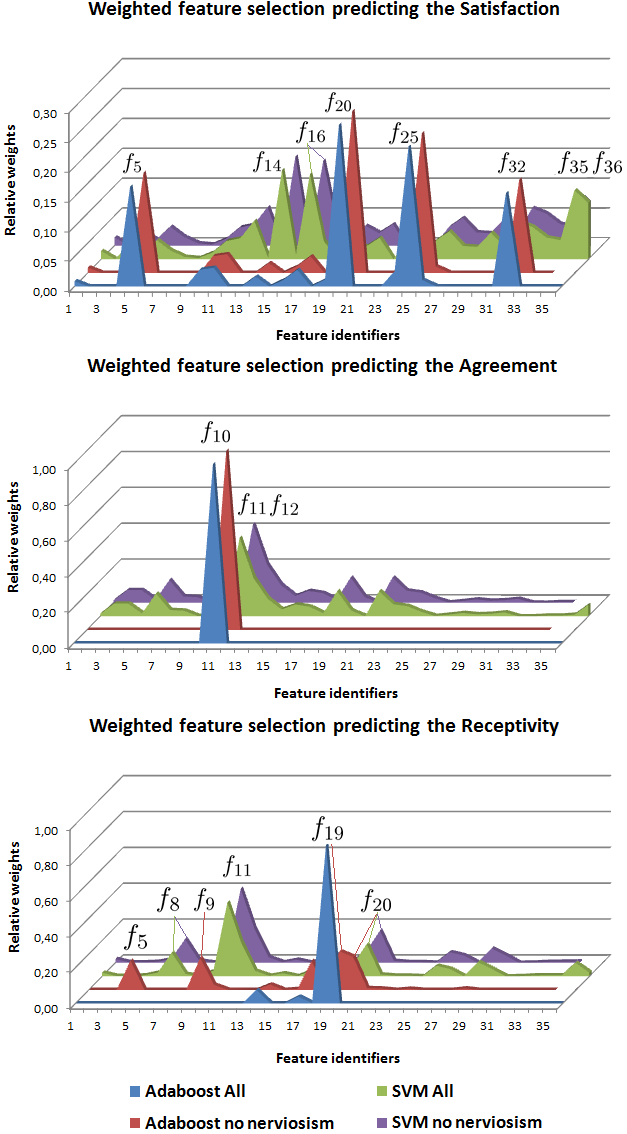}
    \caption{Weighted feature selection when using AdaBoost and SVM for the grouping response cases presenting the highest accuracy when predicting receptivity, agreement, and satisfaction. Each line represents the relative feature weights assigned by the classifiers within the range $[0,1]$, either employing all features or without the nervousness features $f_{35}$ and $f_{36}$.}
    \label{fig:wfeatures} 
\end{figure}

The result tables show that CF and FF (and even LDA) vary
significantly in their predictions depending on whether the
nervousness features $\{f_{35},f_{36}\}$ are considered or not.
This indicates that the subjective evaluation of the mediator adds
an important weight to the system for half of our classifiers.
Moreover, the variability in performance presented by the
remaining classifiers in relation to these two cases leads us to
analyze the relevance of these features in each case. Thus, we
performed a comparison to identify the most relevant features for
each social response. In this way, we also analyzed the influence
of the nervousness features $\{f_{35},f_{36}\}$ when choosing the
most relevant of the other features. We performed a weighted
feature selection using ~\cite{Escalera:2010:EOC:1756006.1756026}
and~\cite{Chen05combiningsvms} for AdaBoost and SVM, respectively.
For each response (receptivity, agreement, and satisfaction), we
selected those features only for the cases giving the highest
degree of accuracy. See the different plots in
Figure~\ref{fig:wfeatures}. In general, we observe that agitation
features and the mediator's speaking turns are chosen as the most
relevant features when predicting satisfaction. By contrast, the
feature chosen as being most relevant for predicting agreement is
the age of the mediator. In the case of receptivity, the fact of
withholding the nervousness features results in the most
significant changes in the feature selection with respect to the
other responses. However, both hand agitation, gaze, and the
combination of the two are chosen as being the most relevant
features when predicting receptivity. On average, the most
relevant features for all the responses are those involving the
combination of gaze and the agitation of the body regions. This
means that these are the most discriminating behavioral indicators
in the prediction of the degree of receptivity, agreement, and
satisfaction in a conversation such as that maintained in a VOM
process. This feature selection procedure has direct implications
for the observational methodology of non-verbal communication,
since it allows experts in the field of psychology and restorative
justice to focus, in any given conversation, on the most
discriminating behavioral indicators automatically selected
through artificial intelligence.

Finally, we relate the overall training data to the different
ground truth annotations using the epsilon-SVR regression
strategy. In this case, when using the leave-one-out strategy, we
obtain a prediction for each sample within the same range as the
quantified annotations $[1,5]$. In this setting, we also ran the
experiment twice: first, we considered all features, and then left
out the nervousness features. Both cases gave the same average
distances when predicting satisfaction, agreement, and
receptivity, with values of $0.59$, $0.64$, and $0.68$,
respectively. This mean deviation with respect to the ground truth
labels was found to be significant and of great interest to the
team of mediators.

\section{Conclusion}

We have proposed a multi-modal framework for the semi-automatic
analysis of non-verbal communication in VOM sessions. We have
demonstrated the usability of computer vision, signal processing,
and machine learning strategies in conversational processes.
Specifically, we computed a set of features using audio-RGB-depth
data. Then, a heuristic procedure was presented within the
multi-modal feature extraction to improve the continuity in the
detection of positive regions of interest between consecutive
frames. Finally, we have defined an automatic
computation of behavioral indicators used as final features for
learning and classification tasks. We have demonstrated the
applicability of the system for use in the restorative justice
field as a tool for mediators,
obtaining recognition accuracies of 86\%
when predicting satisfaction and 79\% when predicting both
agreement and receptivity, and a high correlation in the
regression analysis.

In future work, we plan to improve the dataset and responses, and
to incorporate new features. In the case of the data, we hope to
capture more samples so as to be able to perform more accurate
predictions, providing continuous ground truth information by
means of intra-mediator estimations. In the case of the
predictions, new data should allow the continuous prediction of
each degree of the behavioral indicators. Moreover, they will
enable us to perform frame-based predictions, analyzing the
evolution of each indicator throughout the VOM process, and to
detect the exact instant (or "click" to use the term employed by
the mediators) when a party accepts the possibility of reaching an
agreement. Finally, we also hope to incorporate emotional state
features, obtained from facial
expressions~\cite{RudovicEtAlPAMI12} and audio data
~\cite{ElAyadi2011572}.


%



%

\ifCLASSOPTIONcaptionsoff
  \newpage
\fi



\bibliographystyle{IEEEtran}
\bibliography{references}
\end{document}